\documentclass{jpsj3}
\usepackage{txfonts}
\usepackage{bm}
\usepackage{color}

\newcommand{\pzt}{PbZr$_{1-x}$Ti$_x$O$_3$} 
\newcommand{\bfco}{BiFe$_{1-x}$Co$_{x}$O$_3$}
\newcommand{\bfo}{BiFeO$_3$}
\newcommand{\bco}{BiCoO$_3$}

\title{First-principles Study on Piezoelectricity and Spontaneous Polarization in Bi(Fe,Co)O$_3$}

\author{Hiroshi Katsumoto$^{1,2}$, Kunihiko Yamauchi$^2$, and Tamio Oguchi$^{2,3}$\thanks{oguchi@sanken.osaka-u.ac.jp}}
\inst{$^1$Graduate School of Engineering Science, Osaka University, Toyonaka 560-8531, Japan \\
$^2$Institute of Scientific and Industrial Research, Osaka University, Ibaraki, Osaka 567-0047, Japan \\
$^3$Center for Spintronics Research Network, Osaka University, Toyonaka 560-8531, Japan} 

\abst{Solid solution \bfco\ shows anti-ferromagnetic order and pyroelectric order, simultaneously. It has been known that BiFe$_{1-x}$Co$_{x}$O$_3$ exhibits a structural phase transition between monoclinic and tetragonal phases as $x$ increases. This kinds of transition is often called morphotoropic phase boundary, which is well known to take place 
in a representative piezoelectric oxide, \pzt. In order to theoretically understand the piezoelectric property in \bfco,
we performed \emph{ab-initio} electronic-structure calculations and studied the structural stability, the magnetic property, and the electronic polarization by means of super-cell approach. 
It turns out that the large electric polarization and the particular pyramidal coordination suppress the response of the electric polarization under strain. 
A way to enhance the piezoelectric effect in \bfco is proposed.}

\begin{document}
\maketitle

\section{Introduction}

Multiferroic materials have attracted an increasing amount of attention over several decades due to the physical interest of the cross-coupling phenomena as well as future industrial applications. Multiferroics indicates more than one ferroic orders at the same time, e.g., the combination of ferroelectric, (anti-)ferromagnetic, and ferroelastic orders.\cite{Ohno_Science1998} A cross-coupling between ferroelectric and magnetic orders may give rise to magnetoelectric (ME) effect, which hopefully enables the electric control of the magnetism and the magnetic control of the ferroelectricity. All the ferroelectric materials show piezoelectric effect to some extent, which is a response of electric polarization to applied  mechanical stress, or a response of mechanical strain by electric field. Multiferroics (ferroelectric + magnetic) materials can also include the piezoelectric effect which may be accompanied by the magnetic response to applied stress (magneto-piezoelectric effect) providing potential applications in spintronics.\cite{Nakajima_Phys.Rev.Lett.2015}

Among the multiferroic materials, BiFeO$_3$ has been considered as the `holy grail' since the antiferromagnetic order and the large spontaneous polarization occurs at room temperature.\cite{Fiebig_NatRevMater2016,Catalan_Adv.Mater.2009} 
BiFeO$_3$ crystallizes in a highly distorted perovskite rhombohedral structure ($R\rm{3}c$) showing the ferroelectric polarization approximately 100 $\mu$C/cm$^2$ along the [111] direction in pseudo-cubic perovskite notation,\cite{Neaton_Phys.Rev.B2005} accompanied by a G-type antiferromagnetic order with N\'eel temperature ($T_{\rm{N}}$) of 640 K.

\begin{figure}[htp]
\centering
\includegraphics[width=0.8\linewidth]{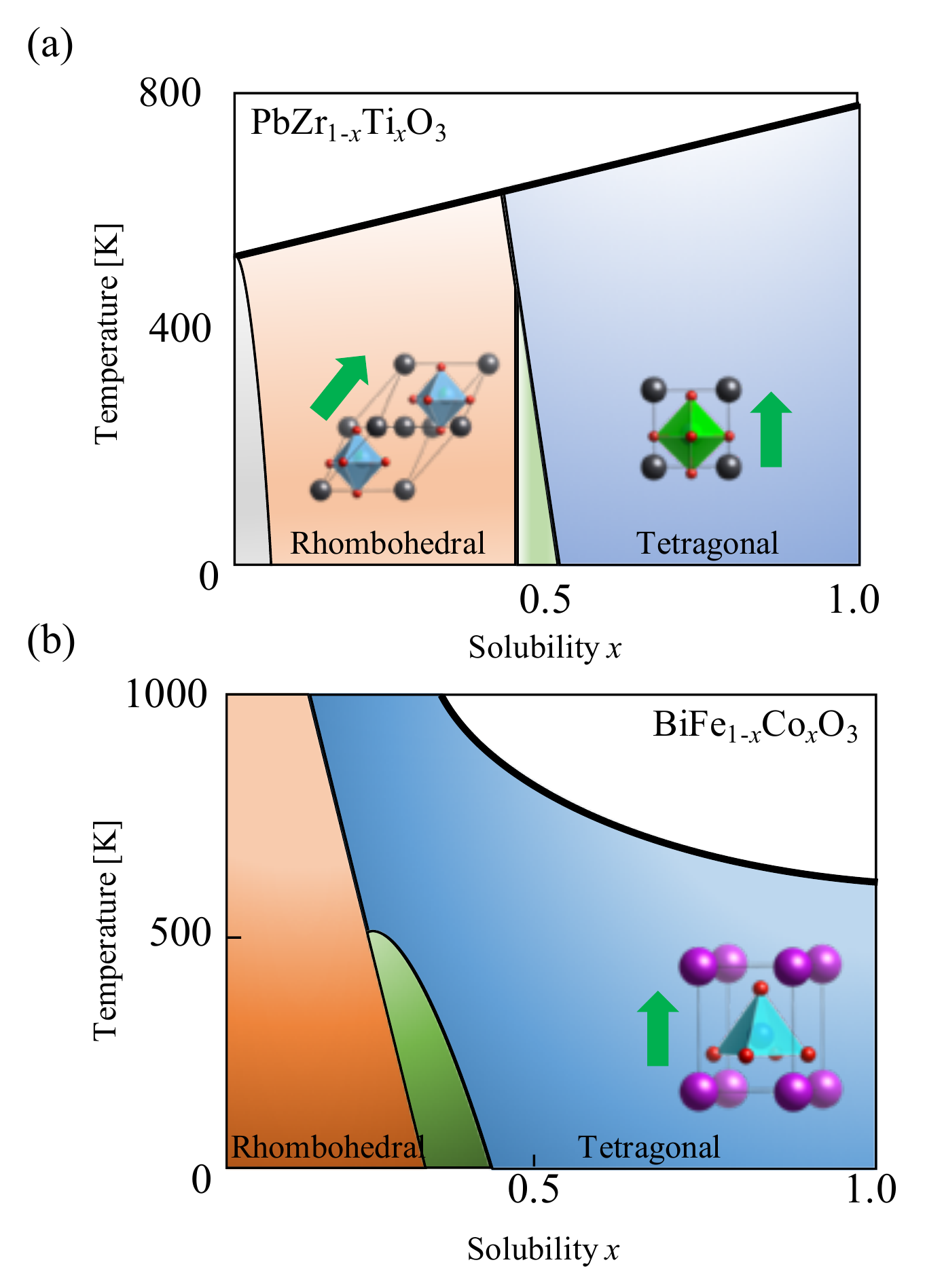}
\caption{\label{fig:phasediagram} (a) Schematic phase diagram of PbZr$_{1-x}$Ti$_{x}$O$_{3}$ based on Ref.~[\citen{Noheda_Phys.Rev.B2000}]. R, M, and T indicates rhombohedral, monoclinic, and tetragonal structure, respectively. Gray part represents orthorhombic structure. (b) Sckematic phase diagram of BiFe$_{1-x}$Co$_{x}$O$_{3}$ based on Ref.~[\citen{Azuma_Jpn.J.Appl.Phys.2008}]. The crystal structures for R-PZT, T-PZT, and T-BFCO are drawn in the corresponding phases. A bold curve represents the Curie temperature.}
\end{figure}

In stark contrast, BiCoO$_3$ crystallizes in a polar tetragonal crystal structure ($P4mm$), quite similar to PbVO$_3$ and shows a C-type antiferromagnetic configuration with $T_{\rm{N}}$ of 470\ K.\cite{Belik_Chem.Mater.2006} The electric spontaneous polarization has been evaluated by a density-functional-theory (DFT) calculation as $P=$179 $\mu$C/cm$^2$.\cite{Uratani_Jpn.J.Appl.Phys.2005} This giant electric polarization is attributed to the strong hybridization among Bi-$p$, O-$p$, and Co-$d$ orbital states, which in turn invokes the high crystal tetragonality ($c$/$a$=1.27, compared to $c$/$a$=1.06 in a prototypical ferroelectric PbTiO$_3$). In the CoO$_5$ pyramidal coordination (see Fig.\ \ref{fig:phasediagram}(b)), the minority-spin electron selectively occupies the $xy$ orbital in Co$^{3+}$ $d$ orbital states.\cite{Uratani_Jpn.J.Appl.Phys.2005} \bco\ represents spin-crossover phenomena with an accompanying structural change at high pressure.\cite{Oka_J.Am.Chem.Soc.2010}

Recently, a series of solid solution \bfco\ (BFCO) has been synthesized in order to tune the electric polarization.\cite{Azuma_Jpn.J.Appl.Phys.2008} Figures \ref{fig:phasediagram}(a) and (b) are schematic phase diagrams of PbZr$_{1-x}$Ti$_x$O$_3$ (PZT) and BFCO, respectively, as a function of solubility $x$. PZT exhibits a high piezoelectric response and is widely used in industry as acoustic sensors and transducers.\cite{Uchino_1997} Around $x$ = 0.5 where PZT shows a maximum piezoelectric effect, a morphotropic phase boundary exists. In Fig.~\ref{fig:phasediagram}(a), the phase diagram of PZT is separated to four parts below the Curie temperature. In the Zr-rich phase ($x<0.5$), PZT is rhombohedral, and on the other hand the crystal structure is tetragonal in the Ti-rich side ($x>0.5$). Between these phases, Noheda $et~al.$ [\citen{Noheda_Phys.Rev.B2000}] found a monoclinic structure ($Cm$) in a vicinity of the morphotropic phase boundary. In this context, Azuma and co-workers expected enhancement in piezoelectricity in the vicinity of the morphotropic phase boundary for BFCO because a monoclinic structure was found in the solid solution BFCO near $x=0.3$ at room temperature,\cite{Oka_Angew.Chem.Int.Ed.2012} indicating a promising candidate of lead-free piezoelectric materials. They analyzed the crystal structure by using X-ray diffraction (XRD) patterns in the phase diagram and searched for a polarization rotation with varying temperatures and ratios of solid solution. The spontaneous polarization could not be measured because of  large leakage current, but was estimated as 117 $\mu \rm{C/cm^2}$ by a point charge model.\cite{Oka_Angew.Chem.Int.Ed.2012} The effective piezoelectric constants has been measured to be $d_{33}\approx$ 60 pm/V in thin film at $x=0.25$.\cite{Shimizu_Adv.Mater.2016}
They intensively studied the electric polarization and magnetic order in the rhombohedral structure of BFCO.\cite{Hojo_Adv.Mater.2018} The enhancement of piezoelectirity is attributed to the presence of a monoclinic phase because the polarization rotates between [001] and [111] in pseudo cubic notation.\cite{Vanderbilt_Phys.Rev.B2001, Wu_Phys.Rev.B2003, Fu_Nature2000} However, this scenario is not sufficient for BFCO with respect to the polarization rotation. Microscopic investigations may lead to deep insights for the piezoelectricity. In this paper, we calculate the electric polarization and the piezoelectric constants in monoclinic BFCO by using a first-principles DFT approach and investigate the microscopic behavior of ions to mechanical strain.

\section{\label{sec:level2}Theoretical Methods}
We performed DFT calculations using Vienna \emph{ab initio} Simulation Package (\textsc{VASP}) within the framework of the generalized gradient approximation adopting the Perdew-Burke-Enzerhof functional (GGA-PBE), while the electron correlation was taken into account by using GGA+$U$ potential (with $U$ = 3 eV for transition metal ions\cite{Walsh_Phys.Rev.B2007}, to optimize the crystal structure and evaluate the electric polarization and the Born effective charge. Spin-orbit coupling was not taken into account.

The  experimental SXRD study has shown that the crystal structure of BiFe$_{2/3}$Co$_{1/3}$O$_3$ is monoclinic $Cm$ structure.\cite{Oka_Angew.Chem.Int.Ed.2012} While Fe and Co atoms are chemically disordered at the perovskite $B$ site, in the present calculations, these atoms are ordered with several structural ordered patterns. The structural stability was examined as varying solubility: $x$ = 0, 1/4, 1/3, 1/2, 3/4, 1 in the monoclinic and tetragonal structures, whereas in the case of rhombohedral structure the stability is examined for $x$ = 0, 1/3, 2/3, 1. 

The spontaneous polarization is calculated by using the Berry-phase method.\cite{Resta_Ferroelectrics1992,Vanderbilt_Phys.Rev.B1993} The calculation of the polarization is not straightforward since the centrosymmetric (paraelectric) structure of BiCoO$_3$ leads to a metallic state. This is due to the degeneracy of Co 3$d$-$t_{2g}$ states that is lifted when the polar distortion invokes Jahn-Teller splitting. A computational technique to make an adiabatic path between ferroelecric and antiferroelectric structures avoiding the metallic state is given in appendix~\ref{app:1}. 

Piezoelectric tensor is calculated in a finite strain response scheme.\cite{Saghi-Szabo_Phys.Rev.Lett.1998} The modulated polarization of ferroelectric system  under applied strain is defined as $P^{T}_{i}=P^{S}_{i}+\sum_{\nu}e_{i\nu}\epsilon_{\nu}$ where $P^{S}$ is original spontaneous polarization and subscripts $i$ and $\nu$ denote direction of polarization and Voigt notation, respectively. The piezoelectric coefficients are expressed as
\begin{equation}
e_{iv}=\tilde{e}^{(0)}_{iv}+\sum_{\alpha,j}\frac{ea_i}{\Omega}Z^*_{\alpha,ij}\frac{\partial u_{\alpha,j}}{\partial \epsilon_{v}}
\label{eq:one}
\end{equation}
where $\Omega$ is the volume of a unit cell, $a_i$ is the lattice parameter, $\alpha$ is an ionic index, $u$ is the internal coordinates, $Z^*$ is the Born effective charge and $\epsilon_{v}$ is the strain tensor element. In Eq.~(\ref{eq:one}), the first term is the clamped-ion term and the second term is the internal strain term. The former is a lattice contribution to the piezoelectricity and the latter comes from  relaxation of the ionic coordinates, which is further decomposed into ionic contributions. For example,  $e_{13}$ is defined as a  change of polarization along the $x$ axis by the strain along the $z$ axis. The clamped-ion term is needed to be corrected as $e^{(0)}_{ijk}=\tilde{e}^{(0)}_{ijk}+\delta_{jk}P^{\textrm{S}}_{i}-\delta_{ij}P^{\textrm{S}}_{k}$, because of an improper contribution from lattice rotation.\cite{Nelson_Phys.Rev.B1976,Vanderbilt_JournalofPhysicsandChemistryofSolids2000} The Born effective charge defined as

\begin{equation}
Z^*_{\alpha,ij}=\frac{\Omega}{ea_i}\frac{\partial P^T_i}{\partial u_{\alpha,j}}
\label{eq:two}
\end{equation}
 satisfies the sum rule $\sum_{\alpha}Z^*_{\alpha}=0$ (as well as the nominal charge). The second term in Eq.~(\ref{eq:one}) is obtained from a product of the Born effective charges and ionic displacement by strain.

\section{\label{sec:level3}Results and Discussions}
\subsection{Structural Analysis}
The tolerance factor is a good indicator for analyzing prefered distortion in the perovskite structure and calculated from ionic radii as 
\begin{equation}
t = \frac{r_{A}+r_{\rm{O^{2-}}}}{\sqrt{2}(r_{B}+r_{\rm{O^{2-}}})}
\label{eq:three}
\end{equation}
 where $r_{A}$ and $r_{B}$ are the ionic radius at the $A$-site and $B$-site cations, respectively, in perovskite oxides.
Table~\ref{tab:tolerancefactor} shows the tolerance factor in the representative perovskite materials and their cation ionic radii. The ionic radius of O$^{-2}$ ($r_{\textrm{O}^{-2}}$) is assumed to be 1.40 \AA .\cite{Shannon_ActaCrystA1976}
SrTiO$_3$ tends to have a cubic perovskite structure according to $t=1$. When the tolerance factor is more than unity, the $B$-site cation has a spacial allowance, that leads to a tetragonal structure with ferroelectlicity. In contrast, when the tolerance factor is less than unity, the $A$-site cation has an allowance with the $B$O$_6$ octahedra rotating and tilting, leading to an antiferroelectric order. Note that the ionic radius of Bi$^{\rm{3+}}$ of 12 coordination was estimated, assuming that ionic radii are proportional to the coordination number in the same valence. BiFeO$_3$ and BiCoO$_3$ have rhombohedral and tetragonal structures, respectively, in the ground state, although their tolerance factors are near unity, indicating that Bi is a key factor in the ferroelectric distortion.
\begin{table}[htp]
\caption{\label{tab:tolerancefactor}%
The tolerance factor $t$ of the representative perovskite materials based on Ref~[\citen{Shannon_ActaCrystA1976}]. $r_{A}$ and $r_{B}$ are the ionic radii of $A$-site and $B$-site cations, respectively.}
\begin{tabular}{lrrr}
\hline
\textrm{}&
\textrm{$r_{A}$ (\AA)}&
\textrm{$r_{B}$ (\AA)}&
\textrm{$t$}\\
\hline
BaTiO$_3$ & 1.61 & 0.61 & 1.06\\
PbTiO$_3$ & 1.49 & 0.61 & 1.02\\
BiCoO$_3$ & 1.45 & 0.61 & 1.00\\
SrTiO$_3$ & 1.44 & 0.61 & 1.00\\
BiFeO$_3$ & 1.45 & 0.65 & 0.98\\
LaCoO$_3$ & 1.36 & 0.61 & 0.97\\
PbZrO$_3$ & 1.49 & 0.72 & 0.96\\
\hline
\end{tabular}
\end{table}

\begin{figure}[htp]
\centering
\includegraphics[width=\linewidth]{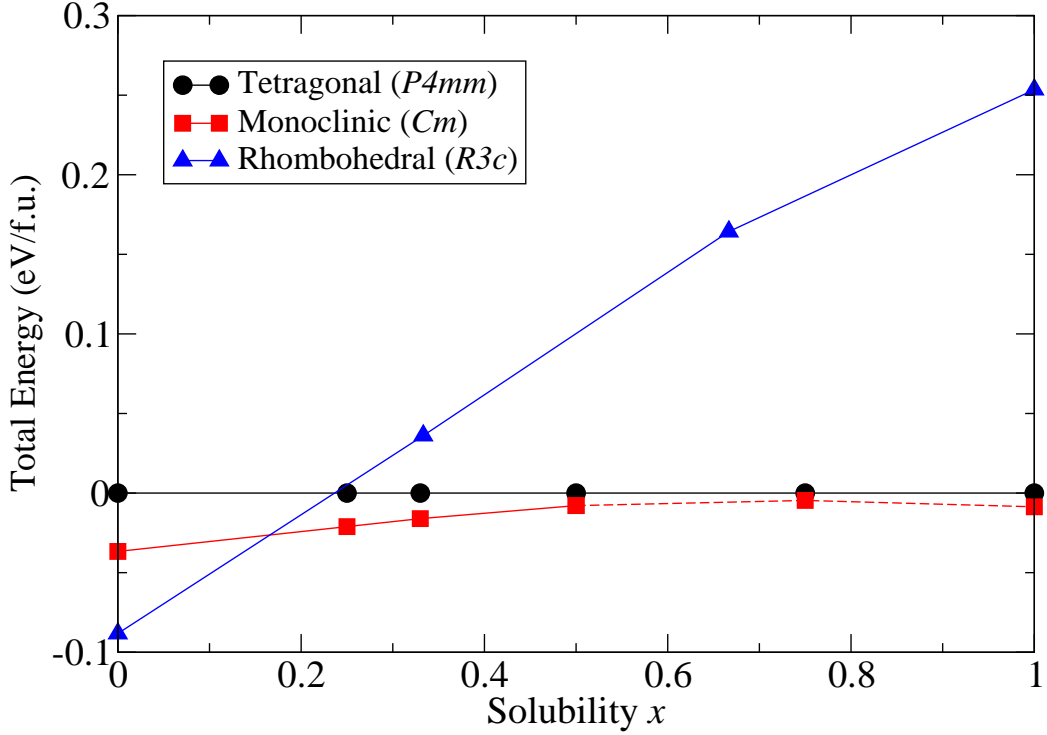}
\caption{\label{fig:sta} 
Total energy as a function of (Fe, Co) solubility $x$ for tetragonal, monoclinic, and rhombohedral structure of \bfco.
 The total energy of tetragonal structure is taken as zero reference for clarity of display.
The closed circle, square, and triangle denote tetragonal, monoclinic, and rhombohedral structure, respectively.}
\end{figure}

The structural stability as varying (Fe, Co) solid solution ratio, $x$, is examined by comparing total energies with tetragonal (space group: $P4mm$), monoclinic ($Cm$), and rhombohedral ($R3c$) structures under the G-type antiferromagnetic configuration. Note that magnetic ground states of \bfo\ and \bco\ in reality are G- and C-type antiferromagnetic, respectively, while the energy difference between these antiferromagnetic orders is prety small compared to the energy difference between different structures. Both the atomic coordinates and lattice parameters are optimized until the atomic force less than 0.005 eV/\AA. As shown in Fig.~\ref{fig:sta}, the overall trend of the stable structure, i.e., rhombohedral structure at $x=0$ (at \bfo), monoclinic structure at $0.2<x<0.5$, and  tetragonal structure at $x=1$ (at \bco), is consistent with experimental results (see Fig. 1(b)). Di\'eguiez and \'I\~niguez reported that the structure change around $x=0.7$ from the formation energy with a DFT calculation.\cite{Dieguez_Phys.Rev.Lett.2011} At $0.5<x$, the monoclinic structure is optimized to be almost identical structure to the tetragonal structure. In PZT the monoclinic structure leads to the enhancement of the piezoelectric coefficients, and therefore the monoclinic structure in \bfco\ may also result in the strong piezoelectric effect.

\subsection{Magnetic Coupling}
BiFeO$_3$ shows the G-type antiferromagnetic order: the Fe$^{3+}$ spin moments are ferromagnetically coupled in the pseudo cubic (111) plane and antiferromagnetically coupled between the adjacent planes. The easy axis of the magnetic moments lies in a plane  perpendicular to the [111] direction. The magnetic symmetry in the AFM phase permits the spins canting in the plane with the cycloidal modulation with a period of 62 nm measured.\cite{Sosnowska_J.Phys.C:SolidStatePhys.1982} In the case of BiCoO$_3$, the Co$^{3+}$ spin moments are ordered in a C-type antiferromagnetic configuration. The magnetic easy axis is in the [001] direction.\cite{Uratani_J.Phys.Soc.Jpn.2009} To our best knowledge, the spin configuration in monoclinic \bfco\ is not known. In order to investigate the magnetic interaction of Fe$^{\rm{3+}}$ and Co$^{\rm{3+}}$ ions in BiFe$_{0.75}$Co$_{0.25}$O$_3$ solid solution, a four-formula-unit monoclinic supercell is adopted. The cell is the minimum setting in a range of solubility in the monoclinic phase ($0.2<x<0.5$). The total energy obtained by DFT calculations was mapped onto a classical Heisenberg Hamiltonian, $H=-\sum_{\langle i,j\rangle}2J_{ij}\bm{s}_{i}\cdot\bm{s}_{j}$, where $\langle i,j\rangle$ denotes pairs of spins at $i$ and $j$ sites and $\bm{s}$ is classical spin. With this cell, the Hamiltonian is described as

\begin{figure}[htp]
\centering
\includegraphics[width=0.7\linewidth]{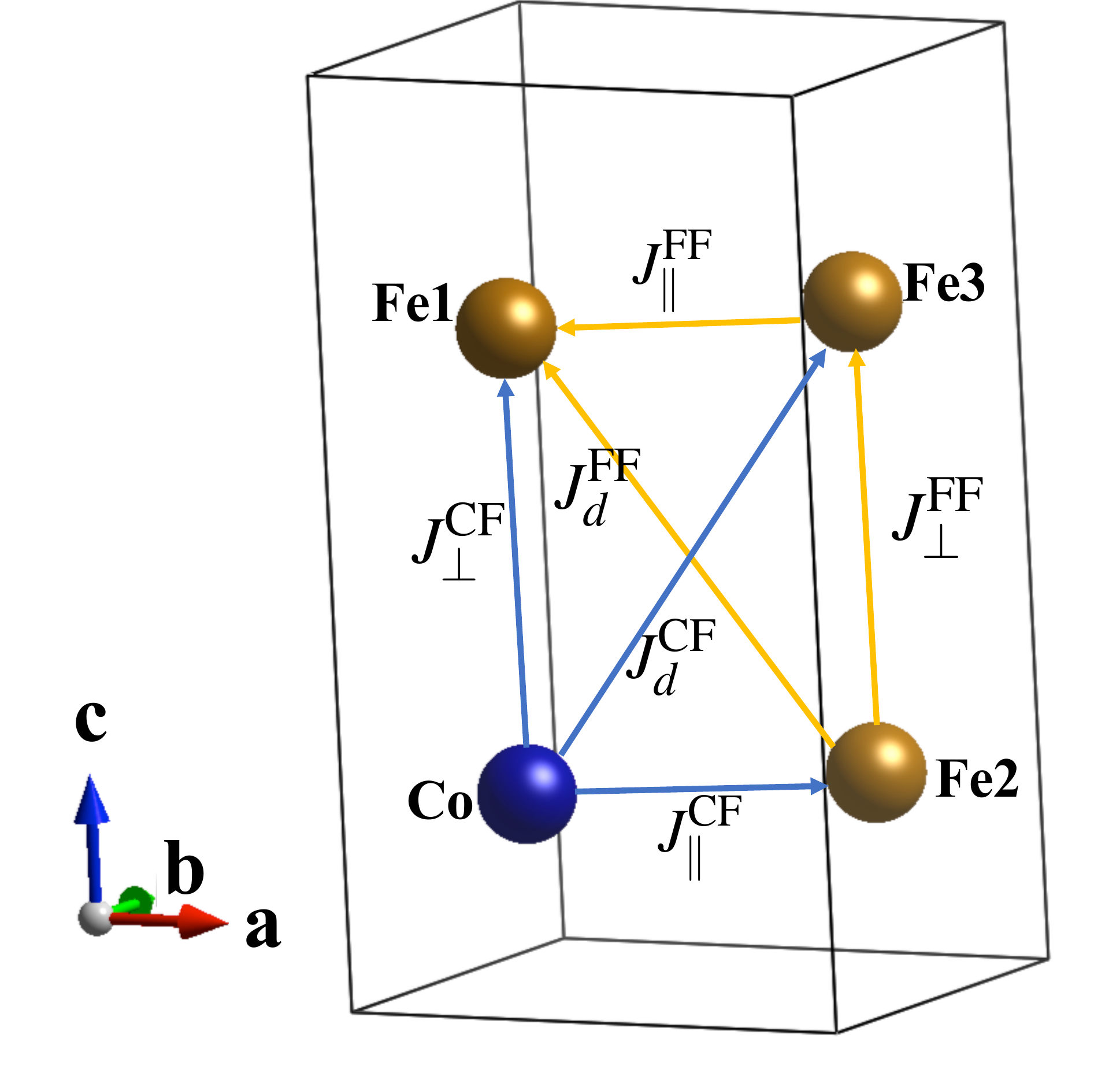}
\caption{\label{fig:cry1} The arrows show spin pairs for the exchange couplings in the model super cell of BiFe$_{0.75}$Co$_{0.25}$O$_3$.}
\end{figure}

\begin{align}
    H&=-4J_{\perp}^{\rm{CF}}\bm{s}_{\rm{Co}}\cdot\bm{s}_{\rm{Fe1}}-8J_{\parallel}^{\rm{CF}}\bm{s}_{\rm{Co}}\cdot\bm{s}_{\rm{Fe2}}-16J_{d}^{\rm{CF}}\bm{s}_{\rm{Co}}\cdot\bm{s}_{\rm{Fe3}} \nonumber\\
     &-4J_{\perp}^{\rm{FF}}\bm{s}_{\rm{Fe1}}\cdot\bm{s}_{\rm{Fe3}}-8J_{\parallel}^{\rm{FF}}\bm{s}_{\rm{Fe2}}\cdot\bm{s}_{\rm{Fe3}}-16J_{d}^{\rm{FF}}\bm{s}_{\rm{Fe1}}\cdot\bm{s}_{\rm{Fe2}}.
\end{align}
where the Heisenberg exchange parameters are displayed in Fig.~\ref{fig:cry1} and coefficients in each term express the coordination number. For example, $J_{\perp}^{\textrm{CF}}$ is an out-of-plane exchange parameter between Fe1 and Co, $J_{\parallel}^{\textrm{CF}}$ is an out-of-plane exchange parameter between Fe2 and Co, and $J_{d}^{\textrm{CF}}$ is an exchange parameter of diagonal pair of Fe3 and Co in the unit cell (see Fig.~\ref{fig:cry1}).

To evaluate the exchange parameters, $\bm{J}=^{T}(J_{\perp}^{\rm{CF}}, J_{\parallel}^{\rm{CF}}, J_{d}^{\rm{CF}}, J_{\perp}^{\rm{FF}}, J_{\parallel}^{\rm{FF}}, J_{d}^{\rm{FF}})$, the energy difference is compared as considering different spin configurations.
 The Heisenberg parameters are calculated by solving a simultaneous equations $A\bm{J}=\Delta\bm{E}$.
 The spin configurations and total energy differences from the ferromagnetic order are listed in Table \ref{tab:spinconfig}. From the energy differences, one can get the Heisenberg exchange parameters to solve Eq.~(\ref{eq:hpara}),
\begin{table}[htp]
\caption{\label{tab:spinconfig}%
Spin configurations and total energy differences from the ferromagnetic order in BiFe$_{0.75}$Co$_{0.25}$O$_3$.} 
\begin{tabular}{ccccc}
\hline
\textrm{Co}&
\textrm{Fe1}&
\textrm{Fe2}&
\textrm{Fe3}&
\textrm{$\Delta E$ (eV)}\\
\hline
$\uparrow$ & $\uparrow$ & $\uparrow$ & $\uparrow$ & 0\\
$\uparrow$ & $\downarrow$ & $\uparrow$ & $\downarrow$ & -0.1339\\
$\uparrow$ & $\downarrow$ & $\downarrow$ & $\uparrow$ & -0.8632\\
$\uparrow$ & $\uparrow$ & $\downarrow$ & $\downarrow$ & -0.9021\\
$\uparrow$ & $\downarrow$ & $\uparrow$ & $\uparrow$ & -0.4287\\
$\uparrow$ & $\uparrow$ & $\uparrow$ & $\downarrow$ & -0.4445\\
$\uparrow$ & $\uparrow$ & $\downarrow$ & $\uparrow$ & -0.5111\\
$\uparrow$ & $\downarrow$ & $\downarrow$ & $\downarrow$ & -0.5129\\
\hline
\end{tabular}
\end{table}

\begin{align}\label{eq:hpara}
A\bm{J}=
\left[
\begin{array}{ccccccc}
8 &  0 & 32 & 8 &  0 & 32 \\
8 & 16 &  0 & 8 & 16 &  0 \\
0 & 16 & 32 & 0 & 16 & 32 \\
0 & 16 &  0 & 0 & 16 & 32 \\
0 &  0 & 32 & 8 & 16 &  0 \\
0 & 16 &  0 & 8 &  0 & 32 \\
8 & 16 & 32 & 0 &  0 &  0 
\end{array}
\right]
\bm{J}=
\left[
\begin{array}{cccccc}
-0.1339 \\
-0.8632 \\
-0.9021 \\
-0.4287 \\
-0.4445 \\
-0.5111 \\
-0.5129
\end{array}
\right]
\end{align}
where this matrix is a coefficient matrix. In order to solve the 7$\times$6 matrix, the least-square solution method is employed as minimizing a sum of square error for each matrix column: $Err=\sum^{7}_{k=1}(\bm{A}_{k}\bm{J}-\Delta E_{k})^{2}$. If all column vectors $\alpha_{k}$ of the coefficient matrix $A$ are linearly independent, $J$ satisfy a following equation, 

\begin{align}
{}^{T}AA\bm{J}={}^{T}A\Delta\bm{E}. 
\end{align}
The calculated Heisenberg exchange parameters in the model super cell of BiFe$_{0.75}$Co$_{0.25}$O$_3$ are listed in Table \ref{tab:Hexparam}. The sum of square error is $6.3\times 10^{-7}$ eV$^{2}$, which is sufficiently small.

\begin{table}[htp]
\caption{\label{tab:Hexparam}%
Heisenberg exchange parameters in the model super cell of BiFe$_{0.75}$Co$_{0.25}$O$_3$.}
\begin{tabular}{lcccccc}
\hline
\textrm{}&
\textrm{$J_{\perp}^{\rm{CF}}$}&
\textrm{$J_{\parallel}^{\rm{CF}}$}&
\textrm{$J_{d}^{\rm{CF}}$}&
\textrm{$J_{\perp}^{\rm{FF}}$}&
\textrm{$J_{\parallel}^{\rm{FF}}$}&
\textrm{$J_{d}^{\rm{FF}}$}\\
\hline
$J$\ [meV] & -2.52 & -27.84 & -1.49 & -3.40 & -23.13 & -1.21\\
\hline
\end{tabular}
\end{table}

In general, the critical temperature is estimated within the mean-field approximation (MFA) for a multisublattice spin system by diagonalizing coupled equations, \cite{Sasioglu_Phys.Rev.B2004}
\begin{equation}\label{eq:four}
\langle\bm{s}_{i}\rangle=\frac{1}{3k_{\rm{B}}T}\sum_{j}2z_{ij}J_{ij}\langle\bm{s}_{j}\rangle   
\end{equation}
where $z_{ij}$ is the coordination number of $J_{ij}$ and $k_{\rm{B}}$ is the Boltzmann constant. Equation (\ref{eq:four}) can be further written down in the form of the eigenvalue matrix problem
\begin{equation}
    (\Theta - T\bm{I})\bm{s} =0,
\end{equation}
where $\Theta_{ij}=(1/3k_{\rm{B}})J_{ij}$, and the eigenvectors express the spin configurations. The N\'eel temperature is 871 K and spin configuration is C-type antiferromagnetic like ferrimagnetic in BiFe$_{0.75}$Co$_{0.25}$O$_3$. In the case of \bco, the N\'eel temperature was 905 K with C-type antiferromagnetic. In the case of $x=0.5$, $T_{\textrm{N}}$ was reported as 400 K in the tetragonal structure with Monte Carlo simulation.\cite{Dieguez_Phys.Rev.Lett.2011}

\subsection{Electric Polarization and Piezoelectricity}
\subsubsection{Piezoelectricity in BiFe$_{2/3}$Co$_{1/3}$O$_{3}$}
\begin{figure}[htp]
\includegraphics[width=\linewidth]{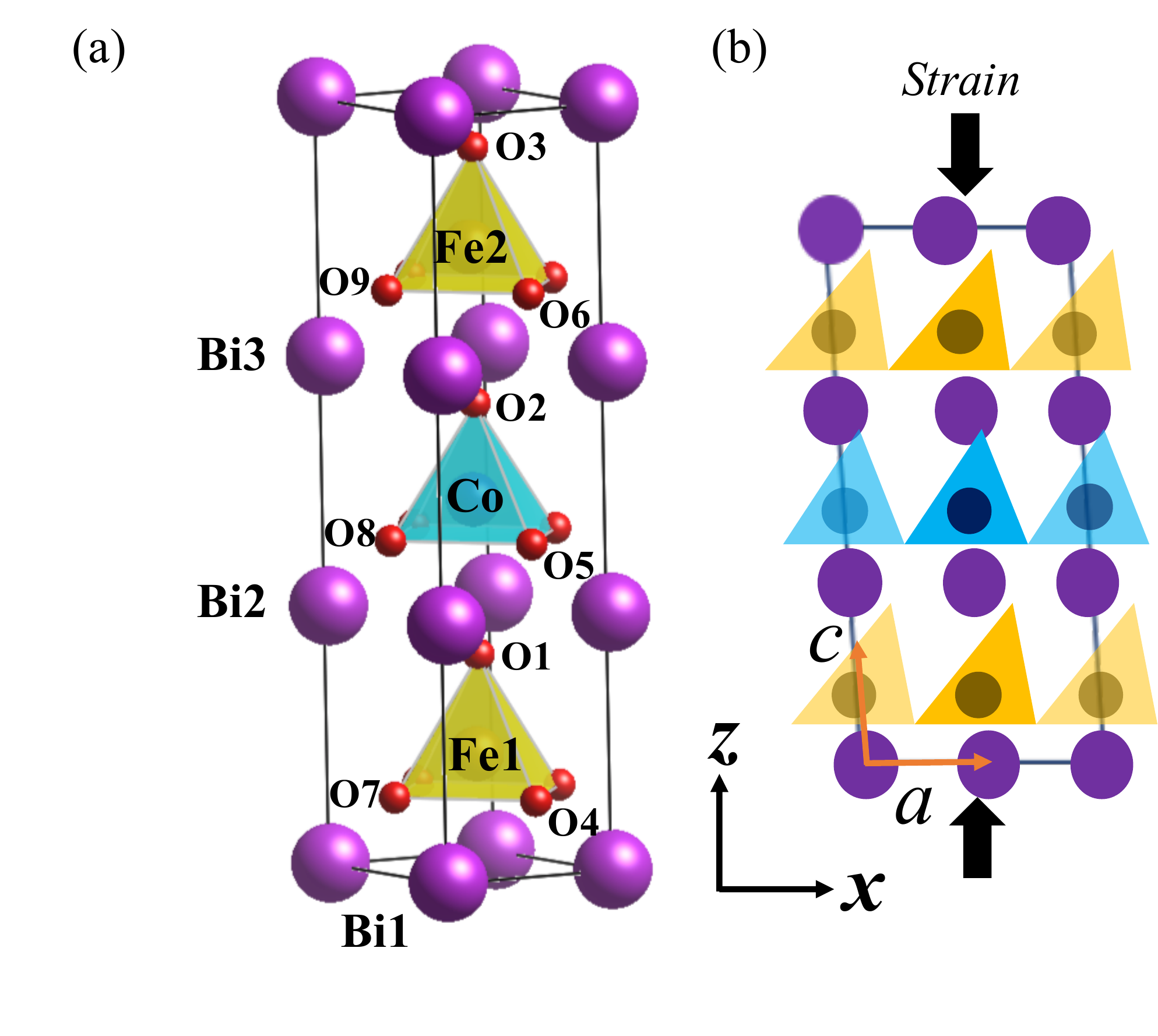}
\caption{\label{fig:cry2} (a) Primitive unit cell of Bi$_{3}$Fe$_{2}$CoO$_{3}$. To host C-type AFM configuration, a $\sqrt{2}\times\sqrt{2}\times 1$ super cell was employed in calculations. (b) Schematic drawing of a mechanical strain and a consequent polarization rotation in which the $x$ axis is parallel to the [110] pseudo cubic direction. The monoclinic distortion is exaggerated. 
}
\end{figure}

In order to calculate the piezoelectric properties in the solid solution \bfco\ with $x=1/3$, a model cell is built, including 30 atoms (see Fig~\ref{fig:cry2}(a)).  The cell is $Cm$ monoclinic structure with a (1$\bar{1}$0) mirror plane. The initial atomic structure is taken from the experimental structure\cite{Oka_Angew.Chem.Int.Ed.2012} reported at $x=0.3$ and is fully relaxed. The lattice parameters are shown in Table~\ref{tab:latticeparam}. Lattice vectors in the monoclinic cell,  $a$ and $b$, are parallel to the pseudo cubic [110] and [$\bar{1}10$] axes, respectively, while $c$ is almost parallel to the [001] direction with a small tilt angle. The polarization components $P_x$ and $P_z$ along the $x$ and $z$ axes of the fully relaxed monoclinic structure are calculated as 155 and $-$35 $\mu \mathrm{C/cm^2}$, respectively. Here, $P_y$ is forbidden by the (1$\bar{1}$0) mirror symmetry. The size of the calculated polarization is consistent with those of a film at $x=0.1$ obtained by experiment.\cite{Hojo_Adv.Mater.2017} The tilting of the polarization direction with respect to the $c$ axis has an important role to enhance the piezoelectricity. When the mechanical strain is applied along the $z$ axis, the total polarization direction may rotate within the mirror plane as shown in Fig.~\ref{fig:cry2}(b). This polarization rotation caused by the simultaneous deformation of oxygen pyramids, significantly decreases $P_z$ and increases $P_x$.

\begin{table}[htp]
\caption{\label{tab:latticeparam}%
The lattice parameters and monoclinic angle of the unit cell in experimental and fully-relaxed calculation. The experimental values$^a$ are taken from reference~[\citen{Oka_Angew.Chem.Int.Ed.2012}]
}
\begin{tabular}{lrrrrrr}
\hline
\textrm{}&
\textrm{}&
\textrm{$a$\ (\AA)}&
\textrm{$b$\ (\AA)}&
\textrm{$c$\ (\AA)}&
\textrm{$\beta$\ (deg.)}&
\textrm{$c/a$}\\
\hline
Exp.$^a$ & BiFe$_{0.7}$Co$_{0.3}$O$_{3}$ & 5.306 & 5.300 & 4.708 & 91.36 & 1.256\\
DFT & BiFe$_{2/3}$Co$_{1/3}$O$_{3}$ & 5.255 & 5.247 & 4.702 & 91.25 & 1.266\\
\hline
\end{tabular}
\end{table}

Piezoelectric $e$-constants are calculated as a combination of a clamped-ion term and an internal-strain term as in Eq.~(\ref{eq:one}). For the first term, the internal coordinates is fixed  and the unit cell is distorted by strain. This term is the contribution from the strained lattice. The second term is calculated by using the Born effective charges $Z^*$ and the slopes of internal coordination with respect to the strain. These slopes are obtained from relaxation calculations for the internal coordinates under strain. The calculated Born effective charges $Z^*$ are listed in Table~\ref{tab:BEC} as obtained by Eq.~(\ref{eq:two}), where each ion is displaced by 0.01 \AA \ ($Z^*$ satisfies the sum rule ($\sum_{\alpha} Z^* =0$)). Difference between the Born effective charges and nominal charges comes from hybridization effect.\cite{Terakura_Proc.Comput.Sci.Workshop2014CSW20142015} For example, Co and Fe ions show larger $Z_{zz}^*$ than $Z_{xx}^*$ and $Z_{yy}^*$ due to the strong hybridization with the apical oxygen ions with the polar distortion while Bi ion  in contrast shows larger $Z_{xx}^*$ and $Z_{yy}^*$ than $Z_{zz}^*$ due to the hybridization with both the apical and side oxygen ions. At the relaxed structure, piezoelectric tensors are calculated as $e_{13}=0.63$ C/m$^2$, $e_{33}=1.67$ C/m$^2$, $e_{11}=3.47$ C/m$^2$, and $e_{31}=1.19$ C/m$^2$. The resulted values are much smaller than what we expected in analogy to PZT ($e_{13}=-33$ C/m$^2$ and $e_{33}=12.6$ C/m$^2$ in PZT\cite{Wu_Phys.Rev.B2003}). In the following, we will discuss this difference in detail. 
\begin{table}[htp]
\caption{\label{tab:BEC}%
The Born effective charges in BiFe$_{2/3}$Co$_{1/3}$O$_3$. From O1 through O3 atoms are apical in the pyramidal structure, and the other oxygens are at the side site.}
\begin{tabular}{lrrr}
\hline
\textrm{Ion}&
\textrm{$Z^{*}_{xx}$}&
\textrm{$Z^{*}_{yy}$}&
\textrm{$Z^{*}_{zz}$}\\
\hline
Co & 2.47 & 2.46 & 3.31\\
Fe1 & 3.05 & 3.04 & 3.99\\
Fe2 & 3.04 & 3.04 & 4.06\\
Bi1 & 5.26 & 5.14 & 3.79\\
Bi2 & 5.30 & 5.25 & 3.61\\
Bi3 & 5.27 & 5.23 & 3.73\\
O1 & -2.55 & -2.54 & -3.40\\
O2 & -2.40 & -2.33 & -3.05\\
O3 & -2.50 & -2.37 & -3.42\\
O4 & -2.91 & -2.89 & -2.20\\
O5 & -2.69 & -2.64 & -1.92\\
O6 & -2.91 & -2.90 & -2.18\\
O7 & -2.91 & -2.89 & -2.21\\
O8 & -2.65 & -2.64 & -1.92\\
O9 & -2.90 & -2.90 & -2.18\\
\hline
Sum & -0.03 & 0.06 & 0.00\\
\hline
\end{tabular}
\end{table}

\begin{figure}[htp]
\includegraphics[width=\linewidth]{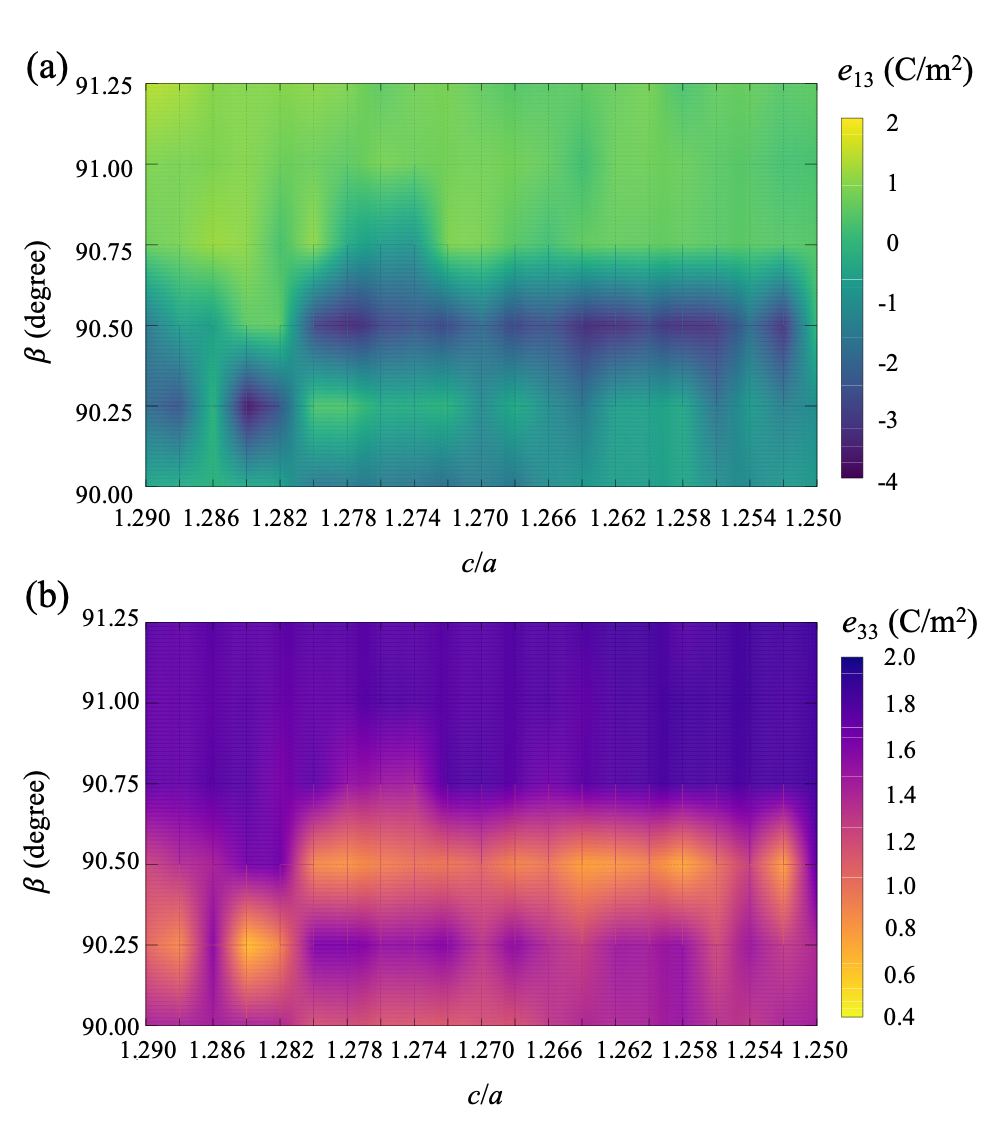}
\caption{\label{fig:piezoe13} Piezoelectric $e$-constants, $e_{13}$ and $e_{33}$ in (a) and (b), respectively, calculated as a function of monoclinic angle ($\beta$) and $c/a$ ratio.}
\end{figure}

In order to investigate how the piezoelectricity develops around the morphotropic phase boundary in \bfco, the monoclinic distortion angle ($\beta$) and $c/a$ ratio are varied at the fixed volume. The $c/a$ ratio is varied  from 1.250 to 1.290 \AA\ with 0.002 \AA\ interval while $\beta$ is varied from 91.25$^\circ$ to 90.00$^\circ$ with 0.25$^\circ$ interval so that the piezoelectric $e$ constants are calculated for total 126 structures. The calculated $e_{13}$ and $e_{33}$ are shown in Fig.~\ref{fig:piezoe13}. The values of $e_{13}$ changes sensitively to $\beta$ (showing large values around $\beta$=90.5$^\circ$) but not to $c/a$. The trends of $e_{13}$ and $e_{33}$ are found to inverse correlation and polarization rotation does not occur when the strain is applied. The maximum magnitude of $e_{13}$ is $-3.77$ C/m$^2$ at $\beta=90.25^{\circ}$ and $c/a=1.284$. On the other hand, the peak value of $e_{33}$ is 1.84 C/m$^{2}$ at $\beta=91.00^{\circ}$ and $c/a=1.258$. It should be emphasized that the enhancement of piezoelectricity occurs within the monoclinic phase, not near the phase boundary of monoclinic and tetragonal structure. In PZT, piezoelectric coefficient, $e_{33}$, is 12.6 C/m$^{2}$ in the vicinity of the morphotropic phase boundary.\cite{Wu_Phys.Rev.B2003} In contrast, $e_{33}$ of the present BFCO shows about seven times as small as that of PZT.

\begin{figure}[htp]
\centering
\includegraphics[width=0.5\linewidth]{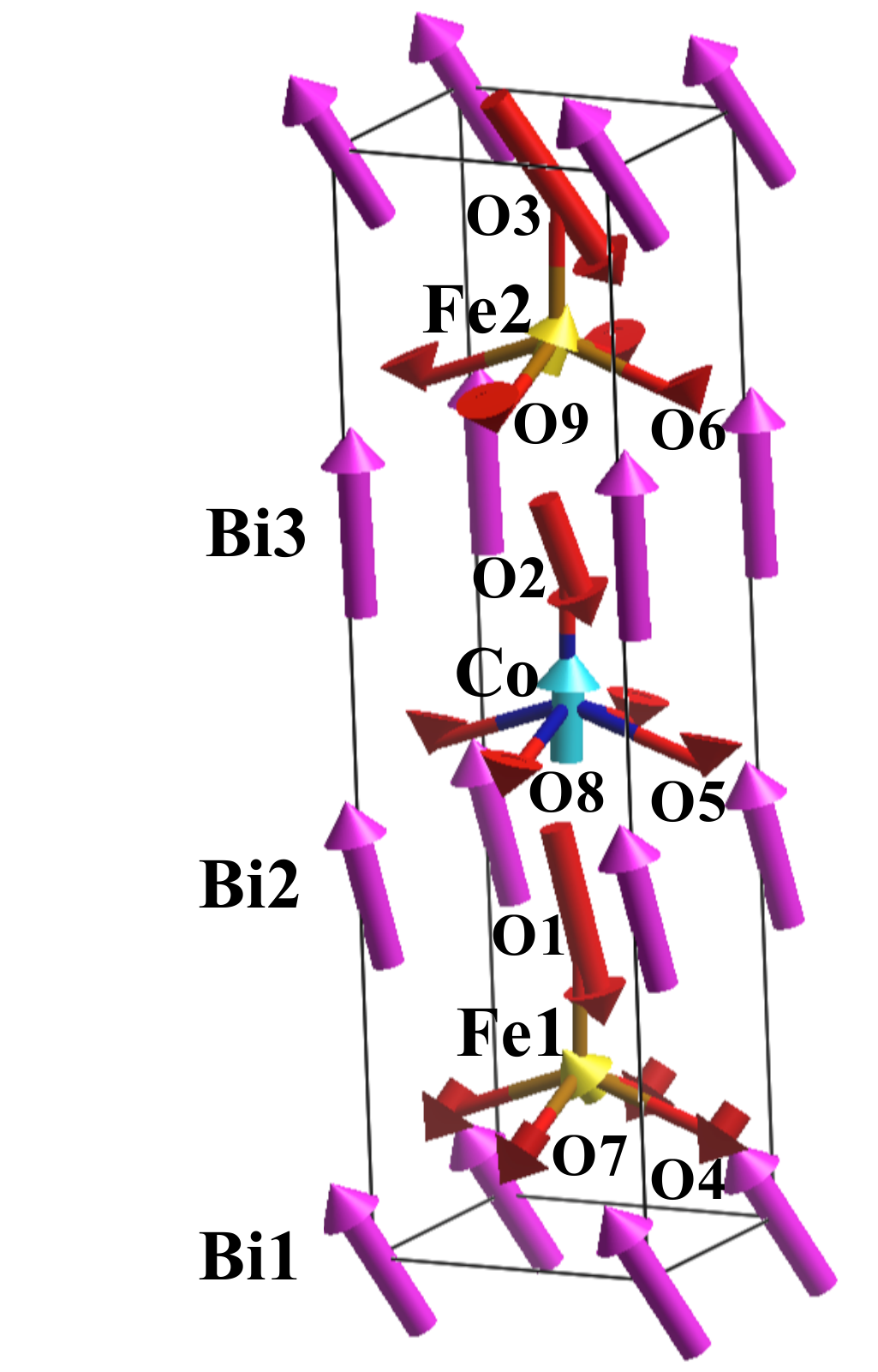}
\caption{\label{fig:slope} The length of arrows shows the displacement magnitude at each atoms. Atomic displacements by $\epsilon_{3}$ strain at $c/a=1.258$ and $\beta=91.00^\circ$ where $e_{33}$ shows the highest value in Bi$_{3}$Fe$_{2}$CoO$_{3}$. The arrangement of atoms are compatible with Fig.~\ref{fig:cry2}(a).}
\end{figure}

In the perovskite piezoelectric oxides, the mechanism of enhancement in the piezoelectric constants is often attributed to the polarization rotation.\cite{Fu_Nature2000, Vanderbilt_Phys.Rev.B2001, Wu_Phys.Rev.B2003} For example,  Wu and Krakauer [\citen{Wu_Phys.Rev.B2003}] have studied the piezoelectricity in PZT with a first-principles calculation and reported that the significant development of polarization along the $x$ axis under strain along the $z$ axis and the polarization rotation increases the piezoelectric constants. In the case of BFCO,  the enhancement of piezoelectricity can be seen in several $c/a$ and $\beta$ points, but rather small. Besides, the enhancements in $e_{33}$ do not have correlation with change in polarization along the $x$ axis. The piezoelectric coefficients are dominantly determined by the internal-stain term in Eq. (\ref{eq:one}), especially the derivative of the ionic displacement with respect to the strain. Figure \ref{fig:slope} shows the ionic displacement by strain at $c/a=1.258$ and $\beta=91.00^{\circ}$, where $e_{33}$ shows the maximum value in our calculations. From Fig.~\ref{fig:slope}, it can be found that the piezoelectricity come from the bismuth and apical oxygen that move along the $c$ direction whereas the transition-metal ions and the side oxygen atoms are not much displaced. 
The ionic behavior comes from the fact that in monoclinic BFCO the apical oxygen atom is not shared by two pyramids so that the apical oxygen ion can move freely but the side oxygen ions and  the transition metal ion in the pyramid are not much affected by the external strain. 
Since the inter-layer interaction of these ionic displacement is weak, the polarization rotates very insensitively under strain. 
This is crucially different from what happens in PZT, where the transition-metal ion in the octahedron cage shows off-centering displacement along the diagonal direction so that  $P_z$ decreases and  $P_z$ increases at the same time. 
Therefore, in BFCO the polarization rotation is not expected to enhance the piezoelectricity. 

To improve such poor piezoelectricity in BFCO, lowering the spontaneous polarization might be considered. 
Phenomenologically, the piezoelectricity can be large with small spontaneous polarization since the piezoelectric coefficients can be represented as a product of dielectric constants and spontaneous polarization, while the dielectric constants decreases more rapidly than the polarization increases.\cite{Budimir_Phys.Rev.B2006}  
Doping La or Y atom at the $A$-site is a possible strategy. According to Cazola {\it et al}.,\cite{Cazorla_Sci.Adv.2017} Bi$_{3/4}$La$_{1/4}$CoO$_{3}$ can be synthesized. 
When the large spontaneous polarization caused by Bi-O hybridization is suppressed, the pyramidal cage will be deformed towards an octahedral cage. 
Then the situation becomes similar to that of PZT, so that the piezoelectricity may be enhanced. 

\subsubsection{Piezoelectric d-constants in BFCO of thin film structures}
The BFCO exhibits two kinds of monoclinic and tetragonal structures in thin film grown on LaAlO$_{3}$ substrate.\cite{Shimizu_Adv.Mater.2016} M$_{C}$-type monoclinic structures are realized in $x$=0--0.10, which lattice tilts toward [010] pseudo cubic direction defined as the $x$ axis with the in-plane electric polarization ($P_{x}$). The second monoclinic structure is M$_{A}$-type monoclinic structures ($x$=0.15--0.40), which are equivalent with the monoclinic structure in bulk. Increasing the solubility ($x\geq 0.40$), the crystal structure becomes tetragonal structure like \bco. The piezoelectric \textit{d}-constants are calculated in M$_{C}$-type monoclinic \bfo\ ($x$=0), M$_{A}$-type monoclinic BiFe$_{0.25}$Co$_{0.75}$O$_{3}$ and BiFe$_{2/3}$Co$_{1/3}$O$_{3}$, and tetragonal BiFe$_{0.5}$Co$_{0.5}$O$_{3}$ shown in Table~\ref{tab:piezod}. The piezoelectric \textit{d}-constants are obtained from a product of piezoelectric \textit{e}-constants and compliance $S$ that is the inverse tensor of the elastic constants ($S=C^{-1}$), for example, $d_{33}=\sum_{i=1}^{6}e_{3i}S_{i3}$ with the Voigt notation. The calculated value for the tetragonal structure shows good agreement with the experimental value. On the other hand, the $d_{33}$ for three monoclinic structures are smaller than the experimental results. It should be noted that the linear response of piezoelectricity was taken into account in our calculation while the experimental values have been obtained from maximum value of non-linear $S$-$E$ curves applying electric fields.\cite{Shimizu_Adv.Mater.2016} Therefor these values are not well comparable. However, our results reproduce the trend of increasing the $d_{33}$ value with decreasing the solubility $x$. The enhancement of piezoelectricity occurs in M$_{C}$-type monoclinic structure. 

\begin{table}[htp]
\caption{\label{tab:piezod}%
Piezoelectric \textit{d}-constants of  $x$=0, 0.25, 0.33, 0.5 in \bfco\ and the experimental values from Ref.~[\citen{Shimizu_Adv.Mater.2016}]}
\begin{tabular}{cccc}
\hline
$x$ & $d_{33}$  & $d_{11}$  & $d_{33}^{\textrm{exp.}}$ \\
\hline
0  & 10.98 & 16.86 & 39.8 \\
0.25 & 17.38 & 69.17  & 52.1--57.5\\
0.33 & 8.54 & 22.22 & -\\
0.5 & 7.84 & 0 & 10.8\\
\hline
\end{tabular}
\end{table}

\begin{figure}[htp]
\includegraphics[width=\linewidth]{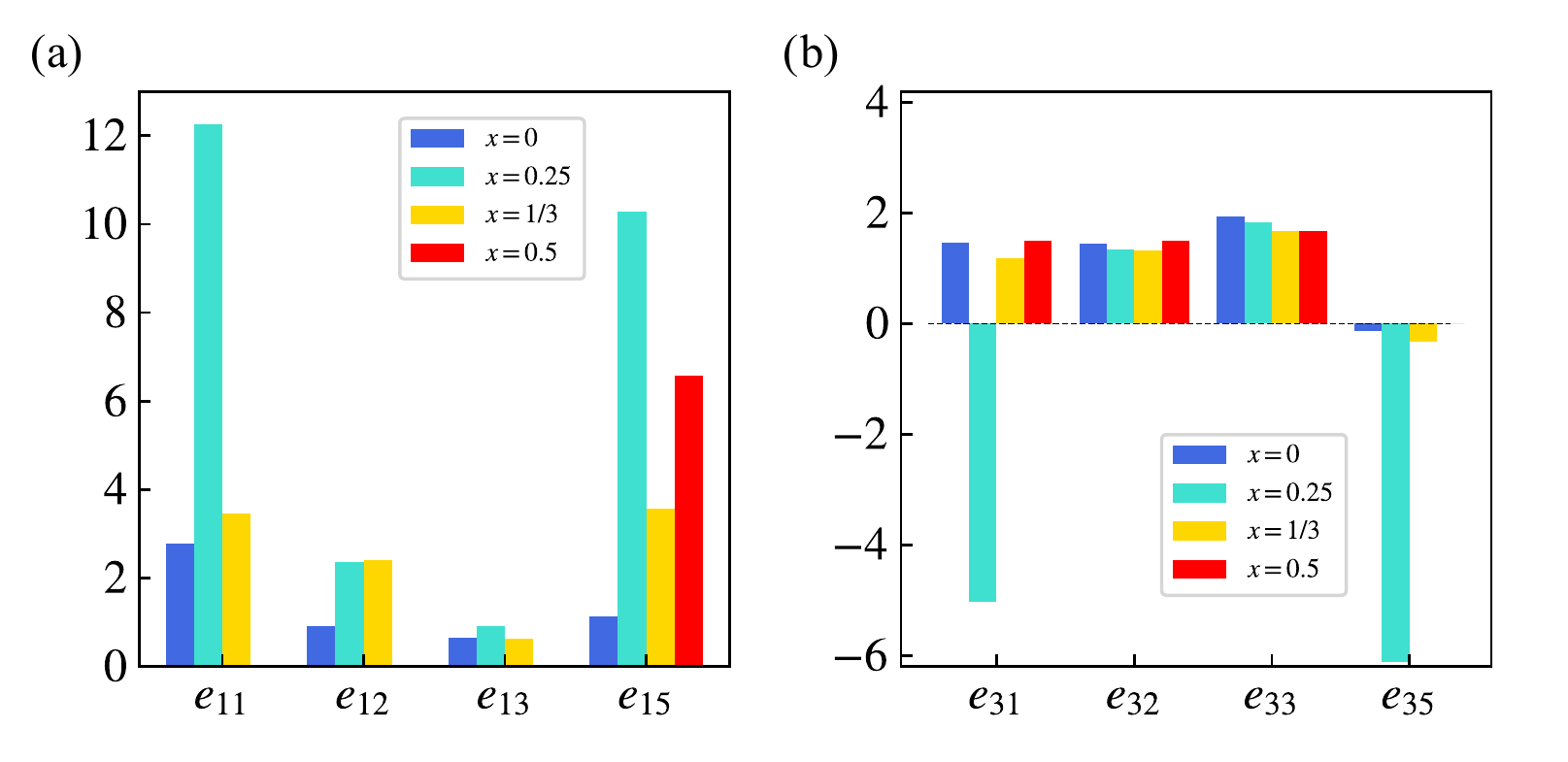}
\caption{\label{fig:piezoe} Piezoelectric $e$-constants in $x=0$, $x=0.25$, $x=1/3$, and $x=0.5$ in \bfco. (a) and (b) show polarization changes along the $x$ and $z$ axes, respectively, by strain.}
\end{figure}

The piezoelectric $e$-constants are enhanced at $x=0.25$ by $\epsilon_{1}$ and $\epsilon_{5}$ strain in Fig.~\ref{fig:piezoe}. The main contribution of $d_{33}$ comes from the $e_{31}$ and $e_{35}$. By contrast, the values of $e_{33}$ are not remarkably changed with the solubility. When the unit cell is strained along the $x$ axis, $z$ component of the electric polarization is decreased. The ionic radii of the Co$^{3+}$ is smaller than that of Fe$^{3+}$ so that the pyramidal structure of Co is considered to be destabilizing at $x$=0.25.The piezoelectric $e$-constants reflect the competition between rhombohedral and tetragonal tendency. At $x=0$, the sign of $e_{31}$ is positive because M$_{C}$-type monoclinic lattice tilts toward [010] pseudo cubic plane where apical and side oxygens are located in the same distorted plane. M$_{A}$-type monoclinic lattice tilts toward [110] pseudo cubic plane so that the apical oxygen can move more because the apical and side oxygens are not in the same strain direction. The sign of $e_{31}$ is changed at $x$ =0.25, as electric polarization $P_{z}$ decreases applying $\epsilon_{1}$ strain. $d_{11}$ mainly comes from $e_{11}$ and $e_{15}$ in Fig.~\ref{fig:piezoe}(b). The BFCO responds greatly to $\epsilon_{1}$ and $\epsilon_{5}$.

\section{\label{sec:level5}Conclusions}
We have evaluated the piezoelectric $e$-constants in monoclinic BFCO at $x=1/3$ as varying $c/a$ ratio and monoclinic distortion angle. The enhancement of piezoelectricity is found in monoclinic structures. The magnitude of $e_{33}$ is rather small since the mechanism of the enhancement of piezoelectricity is different. The polarization rotation in PZT is a plausible explanation of the enhancement of piezoelectricity because the octahedron of Zr and Ti share the apical oxygen. Therefore, in-plane and out-of-plane electric polarizations cooperatively change.  However, in BFCO, bismuth and not shared apical oxygen primarily contribute to the piezoelectricity with displacement of ions by strain. The pyramidal polyhedron make the apical oxygen possible to move independently and to induce the displacement of bismuth. The piezoelectric $d$-constants are calculated in different solubility $x$ and the trend of $d_{33}$ shows that the enhancement of piezoelectricity occur with monoclinic structure. The main factor of the $d_{33}$ is not $e_{33}$, but $e_{31}$ and $e_{35}$. 
The $d_{11}$ is also show the enhancement of piezoelectricity at $x=0.25$ and which is promising material for the lead-free material.

\begin{acknowledgments}
The computations in this study were performed using the facilities of Supercomputer Center at ISSP, University of Tokyo. We would appreciate Dr. Silvia Picozzi and Dr. Paolo Barone for invariable discussion. This work was supported by JSPS Kakenhi, Grants No. 17H02916. 
\end{acknowledgments}

\appendix
\section{\label{app:1}Polarization Calculation in Pyroelectric Materials}
This Appendix is contributed to explain the method of numerical calculation for spontaneous polarization in pyroelectric materials by using DFT calculation. In modern approach, Berry phase method is a good tool for the calculation, but it has uncertainty by quantum polarization. Electric polarization is calculated from paraelectric structure to ferro- (antiferro-) electric one in an adiabatic path. Most materials remain insulating in paraelectric structure but some ferroelectrics, for example BiCoO$_3$, is not insulating in the paraelectric case. Electric polarization should be calculated in an insulating and adiabatic path. The problem is how can we obtain an adiabatic path without a metallic state from paraelectric to ferroelectric phases.

BiCoO$_3$ is an example material, which shows giant spontaneous polarization and high-tetragonality. In the ground state, BiCoO$_3$ is distorted and octahedral CoO$_6$ becomes to pyramidal structure, because of Jahan-Teller effect. Co$^{3+}$ ion is six electron configuration in $d$ orbital, the $d$ states are all occupied in the majority-spin bands, while in the minority-spin bands, only the $xy$ orbital is selectively occupied. In case of paraelectric structure, BiCoO$_3$ shows metallic because of the minority-spin band. 

Our approach is achieved by comparing electric polarization of ferroelectric with antiferroelectic. In this approach, electric polarization is calculated within insulating. Figure~\ref{fig:adiaba} is an adiabatic path from an antiferroelectric to ferroelectric structures. The value of the electric polarization along the $z$ axis is 171 $\mu$C/cm$^{2}$.

\begin{figure}[htp]
\includegraphics[width=\linewidth]{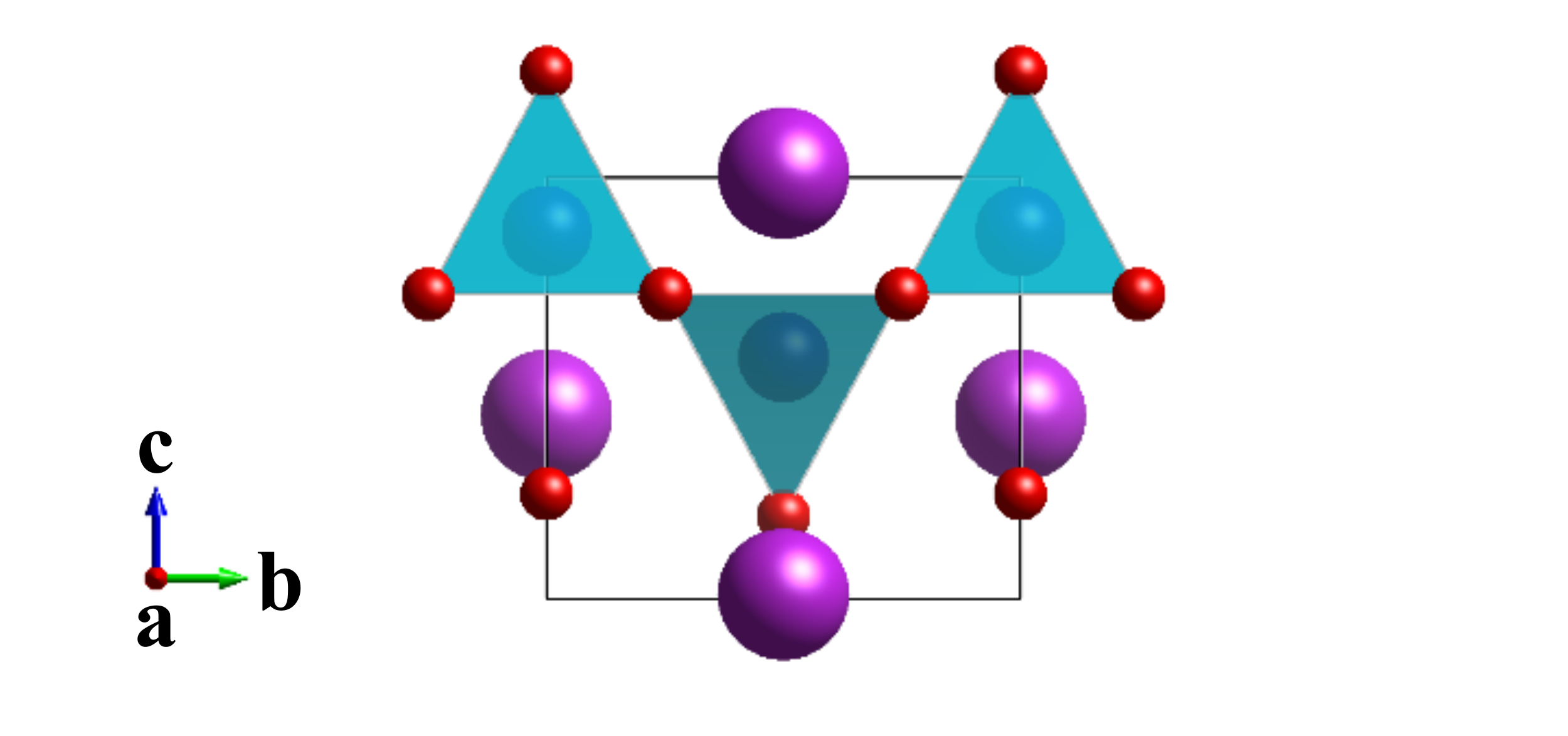}
\caption{\label{fig:cry3} Antiferroelectric structure of BiCoO$_3$ viewed parallel to the $a$ axis. The pyramidal structures are alternating opposite direction with bottom face.}
\end{figure}

\begin{figure}[htp]
\includegraphics[width=\linewidth]{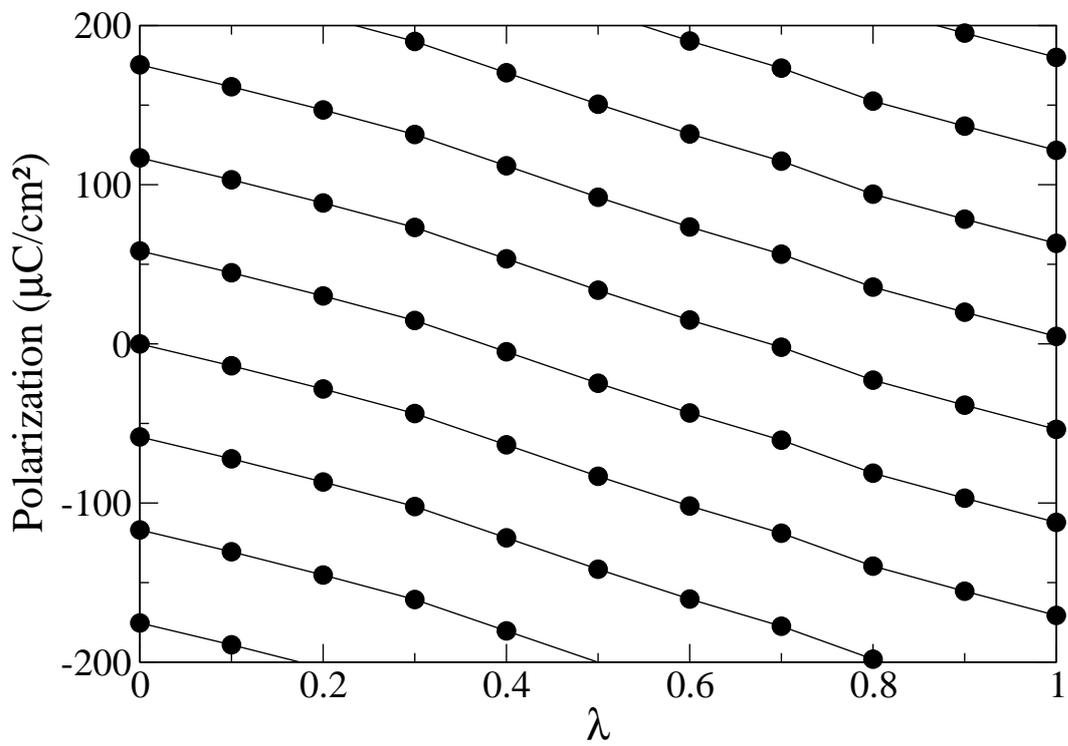}
\caption{\label{fig:adiaba} Calculated polarization with adiabatic path from antiferroelectric to ferroelectric structure, where $\lambda$ scales continuous chages of ionic displacements.}
\end{figure}

\end{document}